\documentclass[aps,twocolumn,letterpaper,longbibliography]{revtex4-1}

\usepackage{amssymb}
\usepackage{amsmath}
\usepackage{graphicx}
\usepackage{extarrows}
\usepackage{url}
\usepackage{varioref}
\usepackage{hyperref}
\usepackage{cleveref}

\usepackage{xcolor}
\usepackage[normalem]{ulem}

\usepackage{verbatim} %Liangchao

\begin{document}

\title{Atomic Bose-Einstein condensate to molecular Bose-Einstein condensate transition} 

% Place the author information here.  Please hand-code the contact
% information and notecalls; do *not* use \footnote commands.  Let the
% author contact information appear immediately below the author names
% as shown.  We would also prefer that you don't change the type-size
% settings shown here.
\author{Zhendong Zhang$^1$}
\author{Liangchao Chen$^2$}
\author{Kaixuan Yao$^1$}
\author{Cheng Chin$^1$}

\affiliation{$^1$James Franck Institute, Enrico Fermi Institute and Department of Physics, University of Chicago, Chicago, Illinois 60637, USA}
\affiliation{$^2$State Key Laboratory of Quantum Optics and Quantum Optics Devices, Institute of Opto-Electronics, Shanxi University, Taiyuan 030006, China}

\begin{abstract}
Preparation of molecular quantum gas promises novel applications including quantum control of chemical reactions, precision measurements, quantum simulation and quantum information processing. Experimental preparation of colder and denser molecular samples, however, is frequently hindered by fast inelastic collisions that heat and deplete the population. Here we report the formation of two-dimensional Bose-Einstein condensates (BECs) of spinning $g-$wave molecules by inducing pairing interactions in an atomic condensate. The trap geometry and the low temperature of the molecules help reducing inelastic loss to ensure thermal equilibrium. We determine the molecular scattering length to be $+220(30)$~Bohr and investigate the unpairing dynamics in the strong coupling regime. Our work confirms the long-sought transition between atomic and molecular condensates, the bosonic analog of the BEC-BCS (Bardeen-Cooper-Schieffer superfluid) crossover in a Fermi gas.
\end{abstract}

% Include the date command, but leave its argument blank.

%\date{}

%%%%%%%%%%%%%%%%% END OF PREAMBLE %%%%%%%%%%%%%%%%

%\usepackage{multicol,lipsum}
%\begin{document} 


%\baselineskip24pt


\maketitle


% For your review copy (i.e., the file you initially send in for
% evaluation), you can use the {figure} environment and the
% \includegraphics command to stream your figures into the text, placing
% all figures at the end.  For the final, revised manuscript for
% acceptance and production, however, PostScript or other graphics
% should not be streamed into your compliled file.  Instead, set
% captions as simple paragraphs (with a \noindent tag), setting them
% off from the rest of the text with a \clearpage as shown  below, and
% submit figures as separate files according to the Art Department's
% instructions.

%\clearpage
%\sout{
%Regular pattern formations arise in nature everywhere and the appearance and evolution of these phenomena have been a focus of recent research activity across several disciplines such as fluids, solid-state physics, nonlinear optics, chemistry and biology\cite{Cross1993}. 
%}
%\textcolor{red}{
%Pattern formation is ubiquitous in nature. ....
%}

% \begin{itemize}
%     \item 
%     General introduction: biology etc
%     \item
%     Classical physics: Rayleigh-Benard convection, Faraday wave $\rightarrow$ Swift-Hohenberg equation
%     \item
%     Equation
%     \item
%     Quantum physics: Abrikosov vortex (SC), cavity polaritons 
% \end{itemize} 

Because of their rich energy structure, cold molecules hold promises to advance quantum engineering and quantum chemistry \cite{Bohn2017, Carr2009, Quemener2012}; a wide variety of platforms are developed to trap and cool cold molecules \cite{Carr2009}. The same rich energy structure, however, also causes complex reactive collisions that obstruct experimental attempts to cool molecules toward quantum degeneracy. 

One successful strategy to prepare molecular quantum gas is to begin with an atomic quantum gas, and then pair the atoms into molecules \cite{Julienne2006}. A prominent example is the pairing of atoms in a two-component Fermi gas, which opens the door to exciting research on the BEC-BCS crossover \cite{Chen2005, Giorgini2008}. Recently, degenerate Fermi gas of ground state KRb molecules is observed based on quantum mixtures of Rb and K atoms \cite{Marco2019}. In these examples, molecules gain collisional stability from Fermi statistics and the preparation of molecules in the lowest rovibrational state, respectively.

For more generic molecules with many open inelastic channels, inelastic collision rates are difficult to predict and experiments frequently report rates near the unitarity limit, which means that all possible scatterings result in loss~\cite{Mayle2013,Idziaszek2010}. The short lifetime hinders evaporative cooling toward quantum degeneracy. 

\begin{figure}[htb]
     \centering
     \includegraphics[width = 76mm]{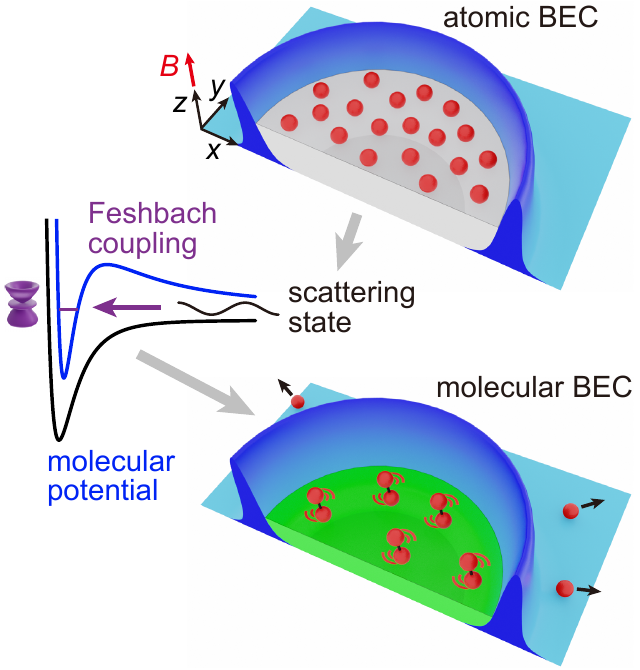}
     \caption{\textbf{Production of $g-$wave molecular condensate.} A uniform Cs BEC (gray) is initially confined in a two-dimensional (2D) optical potential (blue). Cesium atoms (red circles) are paired into molecules in the $g-$wave orbital (purple) through a narrow Feshbach resonance at magnetic field $B_0=19.87$~Gauss in the $z-$direction. The molecules form a molecular BEC (green) in the same optical trap, while the remaining atoms are expelled from the trap.}
     \label{fig:Fig1}
 \end{figure}

Here we report the observation of BECs of Cs$_2$ molecules in a high vibrational and rotational state, see Fig.~\ref{fig:Fig1}. The molecules are produced by pairing Bose-condensed cesium atoms in a two-dimensional, flat-bottomed trap near a narrow $g-$wave Feshbach resonance~\cite{Chin2010}. The trap geometry allows molecules to form with very low temperature and low collision loss such that the molecular condensates emerge in the Berezinskii-Kosterlitz-Thouless (BKT) superfluid regime~\cite{Kruger2007, Tung2010, Hung2011nature,Dalibard2011}. Our experiment opens exciting possibilities to investigate pairing and unpairing dynamics in a bosonic many-body system, described by the interaction Hamiltonian \cite{Romans2004,Leo2004,Timmermans2001,Duine2004}

\[ H_{int}=g (\hat{a}_m^\dagger\hat{a}\hat{a} +\hat{a}_m\hat{a}^\dagger\hat{a}^\dagger),\]

\noindent where $\hat{a}_m$ and $\hat{a}$ are the annihilation operators of a molecule and an atom, respectively, and $g$ is the coupling constant. Pairing in an atomic BEC is expected to induce a quantum phase transition into a molecular BEC \cite{Romans2004}.

\begin{figure*}[htb]
    \centering
    \includegraphics[width = 152mm]{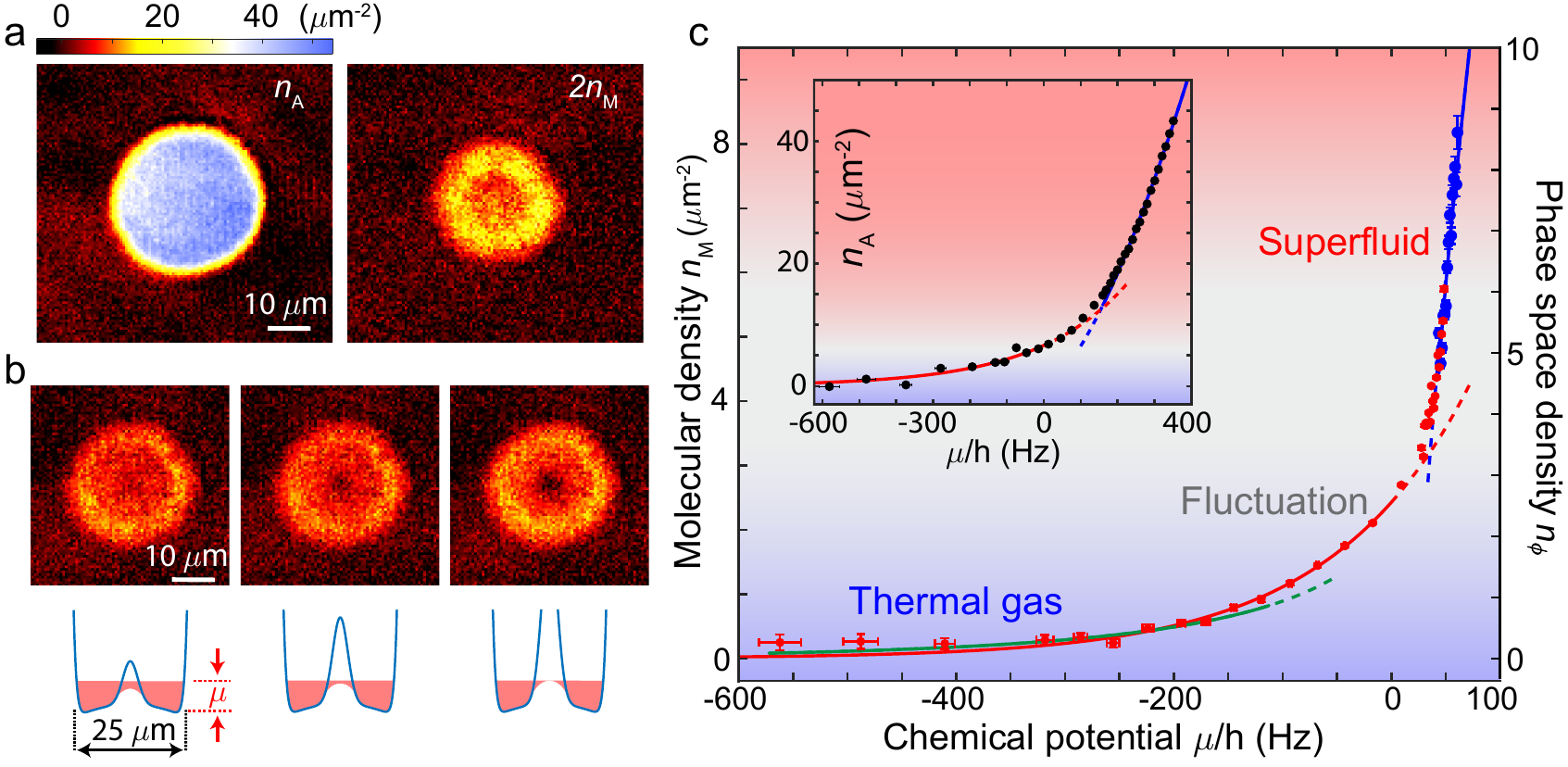}
    \caption{\textbf{Equation of state of molecular gases.} \textbf{a}, \textit{In situ} images of atomic BEC (left) and molecular BEC (right), both at $B=19.2$~G, in the dipole trap. Atoms are paired into molecules near the $g-$wave Feshbach resonance, see text. \textbf{b}, Molecular density response to optical potential. A circular repulsive barrier with a radius of 4~$\mu$m is raised at the center of the trap with a barrier height of $h\times 83$ (left), $h\times165$ (middle) and $h\times330$~Hz (right). Molecular density response determines the equation of state for small and negative chemical potential~\cite{Supplement}. The total external potential is sketched below. \textbf{c}, Equation of state of atomic and molecular BEC. Two dimensional phase space density $n_{\phi}=n_\mathrm{M}\lambda_{\mathrm{dB}}^2$ of molecules are derived from the optical barrier (red) and density profile (blue) measurements~\cite{Supplement}, where $\lambda_{\mathrm{dB}}$ is the molecular de-Broglie wavelength. Background color shows the 2D gas in the thermal ($n_{\phi}\leq 2$, blue), fluctuation ($2<n_{\phi}<n_c$, grey) and BKT superfluid ($n_{\phi}>n_c$, red) regimes, where the superfluid and BEC critical phase space density is $n_c = 6.5$(exp) and  7.5(theory), see text. Green and blue lines are fits in the thermal and superfluid regimes for a 2D Bose gas~\cite{Svistunov2002,Supplement}, respectively. The red line is a fit based on classical gas. Inset shows identical measurement on atomic condensates with fits in the thermal (red) and BEC (blue) regimes. Error bars represent 1-$\sigma$ standard deviation.}
    \label{fig:Fig2}
\end{figure*}

Our experiment starts with a BEC of $6\times10^4$ cesium atoms in a uniform optical trap. The radial confinement on the $x-y$ plane comes from a circular, flat-bottomed optical potential \cite{Clark2017}. The sample is vertically confined to a single site of an optical lattice with trap frequency $\omega_z=2\pi \times 400$~Hz and $1/e$ radius 0.4~$\mu$m. The atomic scattering length is $127~a_{\mathrm{0}}$ at $B=19.2$~G, the global chemical potential is $\mu = h\times 365$~Hz and the temperature is $11(2)~$nK, where $h = 2\pi\hbar$ is the Planck constant and $a_{\mathrm{0}}$ is the Bohr radius.

We create Cs$_2$ molecules by ramping the magnetic field across a closed-channel dominanted Feshbach resonance at $B_0 = 19.87$~G ~\cite{Herbig2003}. This resonance has a small width $\Delta B = 11$~mG~\cite{Innsbruck2018} and couples two scattering atoms into a weakly-bound molecule with a large orbital angular momentum $l=4\hbar$ and projection along the molecular axis $m_l=2\hbar$. The molecules are short-ranged with the size given by the van der Waals length $R_\mathrm{vdW}=5.3~\mathrm{nm}$ for Cs \cite{Chin2010}. This resonance is chosen due to the superior collisional stability between the molecules.

The ramp is optimized to pair up to $15\%$ of the atoms into molecules with the lowest achievable temperature~\cite{Supplement}. After the molecular formation, residual atoms are optically expelled from the trap and a magnetic field gradient is applied to levitate the molecules \cite{Herbig2003}. To detect the molecules, we dissociate them back to atoms by reversely ramping the field well above the resonance, and perform \textit{in situ} imaging on the atoms, see Fig.~\ref{fig:Fig2}a. 

The produced molecules thus occupy the same trap volume as the atomic cloud. Slightly lower molecular density is observed at the trap center due to a weak magnetic field curvature of 21.5$~\mathrm{G/cm}^2$ on the $x-y$ plane. The field curvature leads to a slightly deeper potential in the rim than the center by $1.1$~nK for the molecules. The appearance of the ring structure in the molecular density profile, see Fig.~\ref{fig:Fig2}a, suggests that the molecules are prepared at a temperature or chemical potential on the order of few nK.

To determine the molecular temperature, we find the conventional time-of-flight method impractical as the molecules expand very slowly within their lifetime. 
%(Because of the low temperature and large size of the sample, the cloud expansion is not obvious within the lifetime.)
Instead we measure the density profile by slowly raising a potential barrier at the trap center over 10~ms and recording the density response, see Fig.~\ref{fig:Fig2}b. With a high potential barrier, the molecules at the center becomes thermal with the density response $\partial n/\partial \mu =n/k_\mathrm{B}T$, where $k_\mathrm{B}$ is the Boltzmann constant. From fitting the data, we determine the molecular temperature to be 10(3)~nK~\cite{Supplement}. The low temperature $k_\mathrm{B}T<\hbar\omega_z$ also suggests that the molecules form a 2D gas.

To probe the phase of the molecules at high densities, we measure the equation of state $n(\mu,T)$ from their \textit{in situ} density distribution ~\cite{Zhou2010}. Precise knowledge of the magnetic anti-trap potential is obtained from identical measurements with atomic condensates~\cite{Supplement}. The molecular density is found to linearly increase with the local chemical potential, consistent with the mean-field expectation $\mu = \hbar^2g_{\mathrm{2D}}n_\mathrm{M}/2m$, where $g_{\mathrm{2D}}= 4 \pi a_\mathrm{M}\sqrt{ 2m \omega_z /h}$ is the 2D coupling constant \cite{Petrov2001}, $n_\mathrm{M}$ is the 2D molecular density and $a_\mathrm{M}$ is the molecular scattering length. Fitting the data with the theoretical prediction including finite temperature contribution~\cite{Svistunov2002,Supplement}, we obtain a temperature of $11(1)$~nK, consistent with the optical barrier measurement.

We combine both measurements to determine the equation of state $n(\mu,T)$ of the molecular gas. In Fig.~\ref{fig:Fig2}c, we present the 2D density $n_{\mathrm{M}}$ as a function of the local chemical potential $\mu$. Notably, the transition from exponential to linear dependence on $\mu$ is the hallmark of the thermal gas to supefluid phase transition. A global fit to the data shows excellent agreement with the theory in the thermal and superfluid limits~\cite{Supplement}. From the fit, we determine the 2D coupling constant $g_{\mathrm{2D}}=0.19(3)$, molecular scattering length $a_\mathrm{M}=+220(30)~a_{\mathrm{0}}$, the peak phase space density $n_{\phi}\approx 9$ and the global chemical potential $\mu_0=h\times61(7)$~Hz.
Repeated experiments in the range of $18.2~\mathrm{G}<B<19.5~\mathrm{G}$ show that $a_\mathrm{M}$ is approximately constant. The peak phase space density exceeds the critical value for the BKT superfluid transition of $n_{c}=6.5$ (exp.)\cite{Hung2011nature} and $7.5$ (theo.) \cite{Svistunov2001} at $g_{\mathrm{2D}}=0.19$. In our flat-bottomed potential, we expect molecular condensation in the superfluid regime \cite{Dalibard2011,Supplement}, and estimate $30\%$ to $50\%$ of the molecules are condensed. 

\begin{figure}[tb]
    \centering
    \includegraphics[width = 84mm]{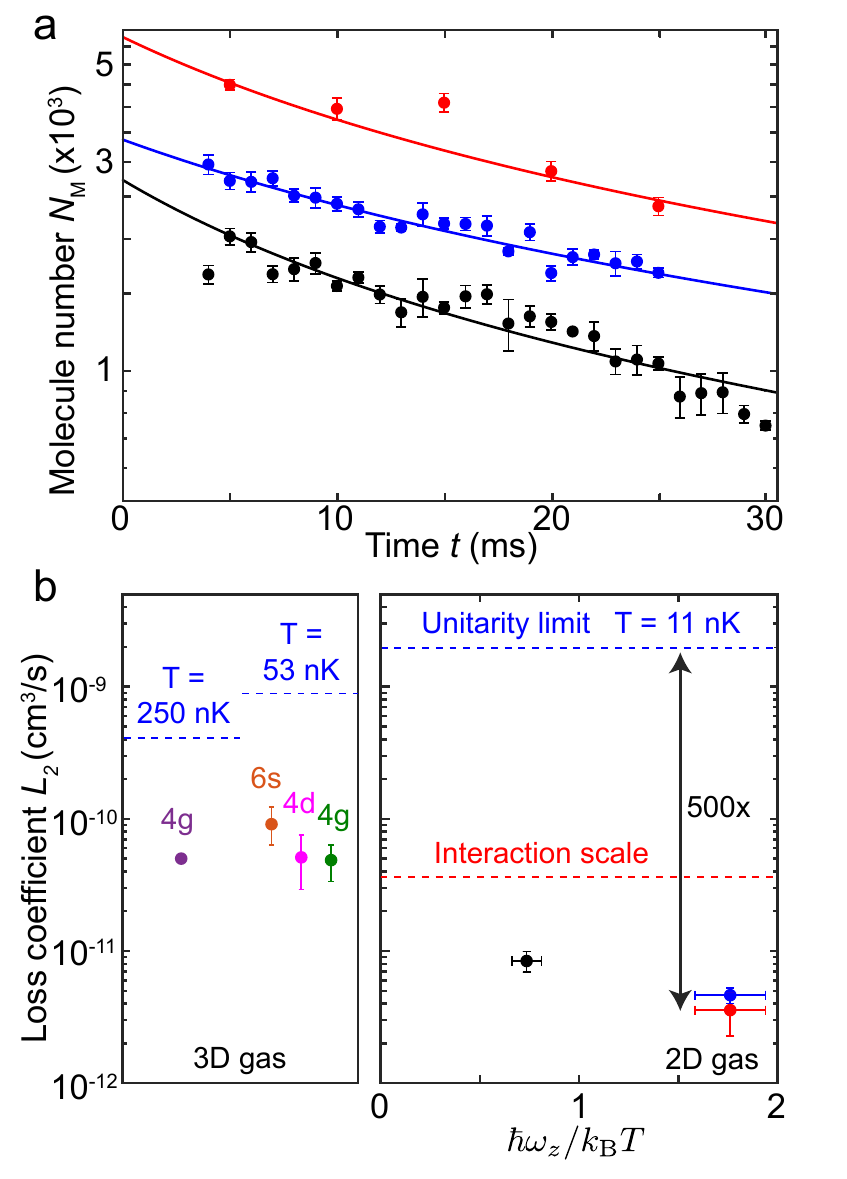}
    \caption{\textbf{Stability of $g-$wave molecular condensate.} \textbf{a}, Decay of total particle number for molecules with vertical trap frequency $\omega_z/2\pi=400$ (red and blue) and $167$~Hz (black) at 19.5~G. The solid lines are fits based on two-body loss rate equation~\cite{Supplement}. \textbf{b}, The extracted loss coefficients $L_2$ in this work (right panel) are compared with former measurements on 3D thermal gases of Cs$_2$ molecules in the $6s$, $4d$ and $4g$ states (left panel)~\cite{Ferlaino2010,Chin2005}. The data points in right panel share the same color code as in panel \textbf{a}. The blue dashed lines indicate the unitarity limits of the two-body loss coefficients. The red dashed line indicates the interaction energy scale $\mu_0/\hbar n_{\mathrm{3D}}$ for the red data point at $10$~ms in panel \textbf{a} with global chemical potential $\mu_0/h=61$~Hz and mean 3D density $n_{\mathrm{3D}}=1.1\times 10^{13}$~cm$^{-3}$. Error bars represent 1-$\sigma$ standard deviation.}
    \label{fig:Fig3}
\end{figure}

We further investigate the lifetime of the molecules. By holding the molecular BEC in the dipole trap with the initial mean density of $n_{\mathrm{3D}}\approx 1\times 10^{13}~\mathrm{cm^{-3}}$, the sample survives for more than $30$~ms. Comparing samples with different densities and in different traps, we conclude that the decays are dominated by two-body collision loss~\cite{Supplement}, see Fig.~\ref{fig:Fig3}a. The average loss coefficient of $L_2= 4\times 10^{-12}$~cm$^3$/s for molecules in the 2D trap with $\omega_z/2\pi = 400$~Hz is significantly lower than previous measurements~\cite{Ferlaino2010,Chin2005}, see Fig.~\ref{fig:Fig3}b . It is also a factor of $500$ below the unitarity limit $U_2 = (4h/2m)\langle k^{-1}\rangle = %\sqrt{16\pi\hbar^4/m^3k_BT} = 
2\times 10^{-9}$cm$^3/$s, where $\langle k^{-1} \rangle$ is the thermal average of the reciprocal molecular scattering wavenumber $k^{-1}$~\cite{Idziaszek2010,Supplement}, and a factor of $10$ below the interaction scale $\mu_0/\hbar n_{\mathrm{3D}}$, see Fig.~\ref{fig:Fig3}b.

The large suppression of inelastic collisions between the highly-excited $g-$wave molecules is remarkable. The comparison in Fig.~\ref{fig:Fig3}b suggests that the collision loss is suppressed at low temperatures and possibly in the 2D regime ~\cite{Idziaszek2015,Micheli2010}. Since the unitarity limited loss scales as $T^{-1/2}$, smaller loss at lower temperature suggests that a larger suppression relative to the unitarity limit can be obtained by reaching down to even lower temperatures. At 10~nK, the loss coefficient we observe is already at the same level as the ground state fermionic molecules reported in Refs.~\cite{Ye2011, Son2020}.

The observed lifetime of 30~ms is sufficient for many elastic collisions between molecules, which occur at the time scale of $\hbar/\mu_0=2.7$~ms, to establish local thermal equilibrium. While the lifetime is insufficient to re-distribute molecules over the entire sample, thermal equilibrium in a (nearly) homogeneous system does not require global transport and can form by local interactions. It is remarkable that the measured temperatures at the trap center and in the rim are in good agreement with the atomic BEC at 11(2)~nK. Our observation suggests that molecules are produced in thermal equilibrium with the atoms. As the result, the molecules are in thermal equilibrium with each other.

\begin{figure*}
    \centering
    \includegraphics[width = 172 mm]{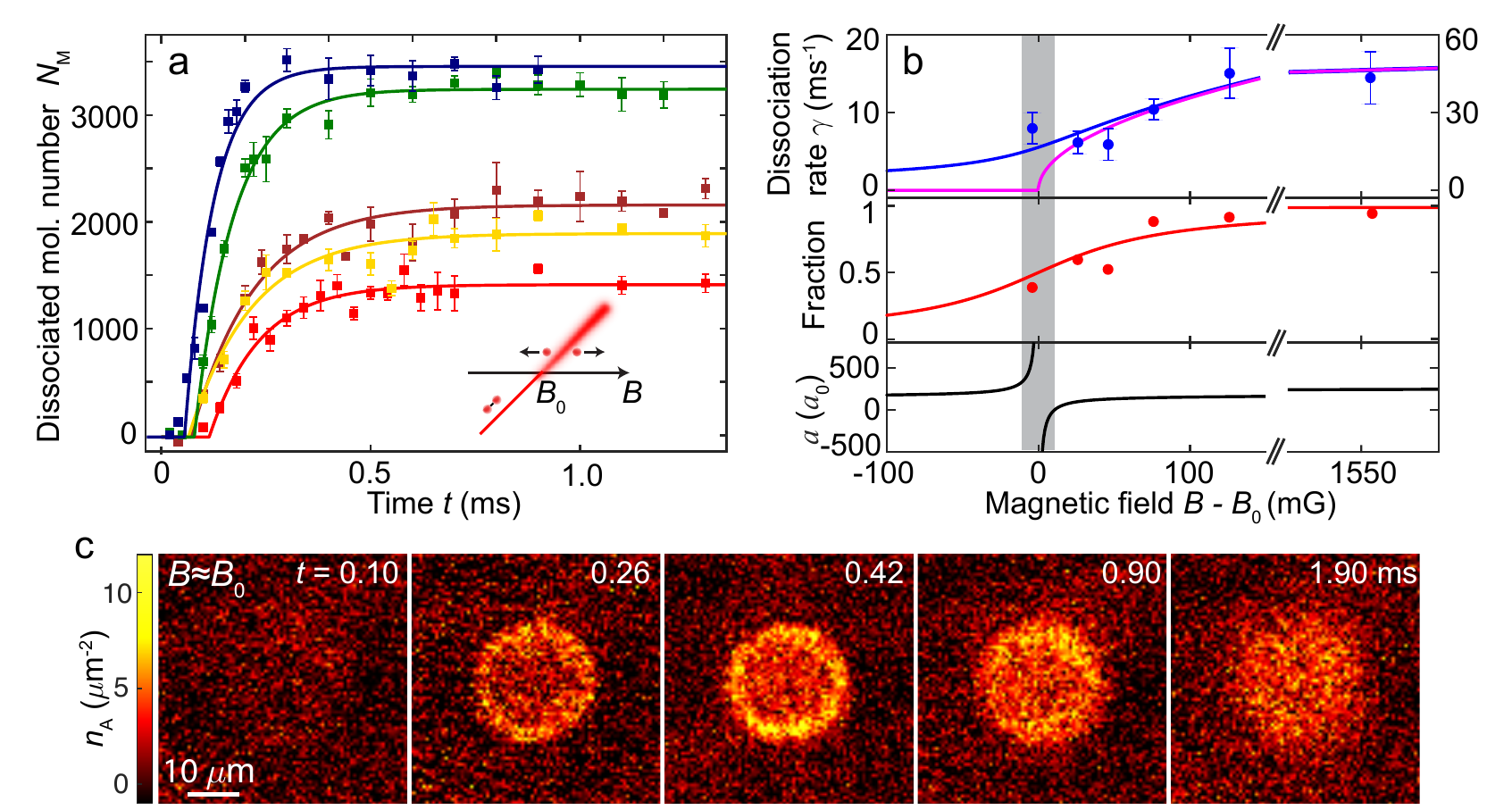}
    \caption{\textbf{Unpairing dynamics in a molecular condensate near the $g-$wave Feshbach resonance at $B_0 = 19.874$~G}. \textbf{a}, 5~ms after the formation of molecular BEC with mean 3D density $n_{\mathrm{3D}} = 9.7\times 10^{12}~\mathrm{cm^{-3}}$, we ramp the magnetic field back to 19.87 (red), 19.90 (brown), 19.92 (yellow), 19.95 (green) and 20.0~G (blue) and image the atoms from the dissociated molecules. The inset illustrates the unpairing process. \textbf{b}, The unpairing rate (upper panel), unpaired fraction (middle panel) extracted from the solid line fits in panel \textbf{a} are compared with the atomic $s$-wave scattering length $a$ (lower panel). The magenta and blue lines are empirical fits based on Fermi's golden rule with the bare and effective density of states, respectively~\cite{Supplement}. The red line is an empirical fit~\cite{Supplement}. The grey shaded area represents the width of the $g-$wave Feshbach resonance. \textbf{c}, \textit{In situ} images of unpaired molecules at $B=19.870(2)$~G near the Feshbach resonance. Error bars represent 1-$\sigma$ standard deviation.}
    \label{fig:Fig4}
\end{figure*}

%To study the transition from molecular BEC to atomic BEC, we ramp the magnetic field to near or above the Feshbach resonance where molecules are dissociated into atoms in the continuum.(change to supplemental)

The molecular superfluid opens a new door to investigate pairing and unpairing in a Bose condensate. A phase transition is expected when unpairing occurs in a molecular BEC \cite{Romans2004,Leo2004,Timmermans2001,Duine2004}. Figure~4 presents our investigation into the unpairing dynamics. After forming the molecular condensate at $B=19.4$~G, we ramp the magnetic field in 0.3~ms near and above the Feshbach resonance with a precision of 2~mG. We monitor the dissociation process by imaging the emerging atoms.

When the field is ramped high above the resonance, the molecules quickly and entirely dissociate. In particular, the dissociation rate follows Fermi's golden rule $\Gamma\propto E^{1/2}$, where $E=\Delta\mu (B-B_0)$ is the molecular energy above the continuum and $\Delta\mu=h\times 770$~kHz/G is the relative magnetic moment \cite{Chin2005}.

Near the Feshbach resonance the system enters the strong coupling regime and the measurement deviates from Fermi's golden rule. Here the measured dissociation energy $\hbar \gamma=\hbar\times 8~\mathrm{ms^{-1}}=k_\mathrm{B}\times$~$61$~nK, see Fig.~\ref{fig:Fig4}b, is much greater than $\mu$ and $T$ of the BEC and much smaller than the Feshbach resonance width $\Delta\mu\Delta B = k_{\mathrm{B}}\times 410~\mathrm{nK}$. The energy is, however, comparable to the universal Fermi energy scale for the molecules $E_F = (\hbar^2/4m)(6\pi^2n_{\mathrm{3D}})^{2/3} = k_\mathrm{B}\times 63~\mathrm{nK}$. This result suggests that the dissociation dynamics near the Feshbach resonance is unitarity-limited ~\cite{Ho2004,Hadzibabic2018}. Finally, we observe about 40\% of the molecules converted back to atoms, and attribute the missing 60\% to inelastic collisions between atoms and molecules in the strong coupling regime.

To conclude, we realize BEC of high-excited, rotating molecules near a narrow Feshbach resonance. The molecules are sufficiently stable at low temperatures to ensure local thermal equilibrium. Unpairing dynamics in molecular condensates is consistent with the universality hypothesis. Our system offers a new platform to study the long-sought atomic BEC-molecular BEC transition, and highlights the fundamental difference between Cooper pairing in a degenerate Fermi gas and bosonic pairing in a BEC.\\

%, smaller than the Feshbach coupling strength $|\Delta\mu\Delta B|=h\times 8.5$~kHz.

%$\gamma = \alpha\Omega\times\sqrt{4\Delta\mu(B-B_0)/\epsilon_{bg}}$~\cite{Ketterle2004}, $\Omega = \Delta\mu\Delta B/\hbar$ is determined by the resonance width $\Delta B$ and the difference of magnetic moments $\Delta\mu$ between a g-wave molecule and two free atoms and $\epsilon_{bg} = \hbar^2/ma_{bg}^2$ where $a_{bg}$ is the background scattering length. 
%Right on resonance $B=B_0$, the dissociation rate reduces to $\gamma=20/$s, see Fig.~\ref{fig:Fig4} (b). 
%This is consistent with Fermi's golden rule that the transition rate $\Gamma$ is proportional to density of states $\rho(\epsilon)$ as $\Gamma = (2\pi/\hbar)|V_{ma}|^2\rho(\epsilon)$, where $\rho(\epsilon) \propto \sqrt{\epsilon}$ in three dimension and the coupling strength $|V_{ma}|^2 \propto a_{bg}\Delta\mu\Delta B$ is determined by the background scattering length $a_{bg}$, difference of magnetic moments $\Delta\mu$ between g-wave molecule and two free atoms and the width of Feshbach resonance $\Delta B$~\cite{Ketterle2004}.

\noindent\textbf{Acknowledgement}

\noindent We thank P. Julienne for helpful discussions and K. Patel for carefully reading the manuscript. This work is supported by National Science Foundation (NSF) grant no. PHY-1511696, the Army Research Office Multidisciplinary Research Initiative under grant W911NF-14-1-0003 and the University of Chicago Materials Research Science and Engineering Center, which is funded by the NSF under grant no. DMR-1420709.

%\bibliographystyle{naturemag}
%\bibliography{natbib}

\clearpage
\widetext
\setcounter{equation}{0}
\setcounter{figure}{0}
\setcounter{table}{0}
\setcounter{page}{1}
\makeatletter
\renewcommand{\theequation}{S\arabic{equation}}
\renewcommand{\thefigure}{S\arabic{figure}}
\renewcommand{\thetable}{S\arabic{table}}

\noindent \textbf{Supplementary Material}\\

\section{Experimental Procedure}

The starting point of our experiment is a BEC of $6\times10^4$ cesium atoms prepared in a disk-shaped dipole trap with a radius of 18~$\mu$m in the $x-y$ horizontal plane. The disk-shaped potential is provided by a digital micromirror device (DMD) which projects 788~nm blue-detuned laser light on the plane with $1~\mu$m resolution. The atoms are loaded into a single layer of the optical lattice in the vertical direction with trap frequency $\omega_z/2\pi = 400$~Hz~\cite{CLthesis}. A magnetic field gradient of 31~G/cm is applied to levitate the atoms and the magnetic field is tuned to 19.9~G. 

We create the molecules with the $g-$wave narrow Feshbach resonance located at 19.87~G based on a procedure similar to Ref.~\cite{Herbig2003}. Since atoms and molecules have different magnetic moments, they tend to separate vertically in the presence of a magnetic field gradient. To better confine both atoms and molecules in the molecular formation phase, we increase the magnetic field gradient to 42.5~G/cm in 2~ms before ramping the magnetic field to 19.76~G in 2~ms which creates molecules. After the formation of molecules, the magnetic field gradient is increased to 50~G/cm in 0.5~ms, which levitates the molecules and overlevitates the atoms.

To remove residual atoms after the molecular formation phase, a resonant light pulse of 20~$\mu$s illuminates and pushes atoms away from the imaging area in 4~ms. Molecules are detected by reversely ramping the magnetic field which dissociates the molecules back to atoms, and the atoms are detected by absorption imaging. The final value of the magnetic field and the hold time are selected to give a reliable image that reflects the distribution of the molecules. In our experiments, we set the final magnetic field to be 20.44~G and do the detection in 0.1~ms after the reverse ramp. We estimate that the atoms expand by 1~$\mu$m during the dissociation process, which is comparable with the imaging resolution of our experimental system.

\section{Characterization of external potential from atomic density profile}
The strong magnetic field gradient for levitating the molecules leads to an additional magnetic anti-trapping potential on the horizontal plane. We also apply a central potential barrier projected from a DMD to measure the density response of the molecules. A precise knowledge of both the magnetic anti-trapping potential and the optical potential barrier are needed in order to extract the equation of state of the molecular gas. 

We load atomic BEC into the same trap as for molecules to calibrate the external potential. Since the magnetic moment and polarizability of the g-wave molecule are accurately known, the trapping potential for molecules can thus be obtained from the trapping potential for atoms.

\begin{figure}
    \centering
    \includegraphics[width=172mm]{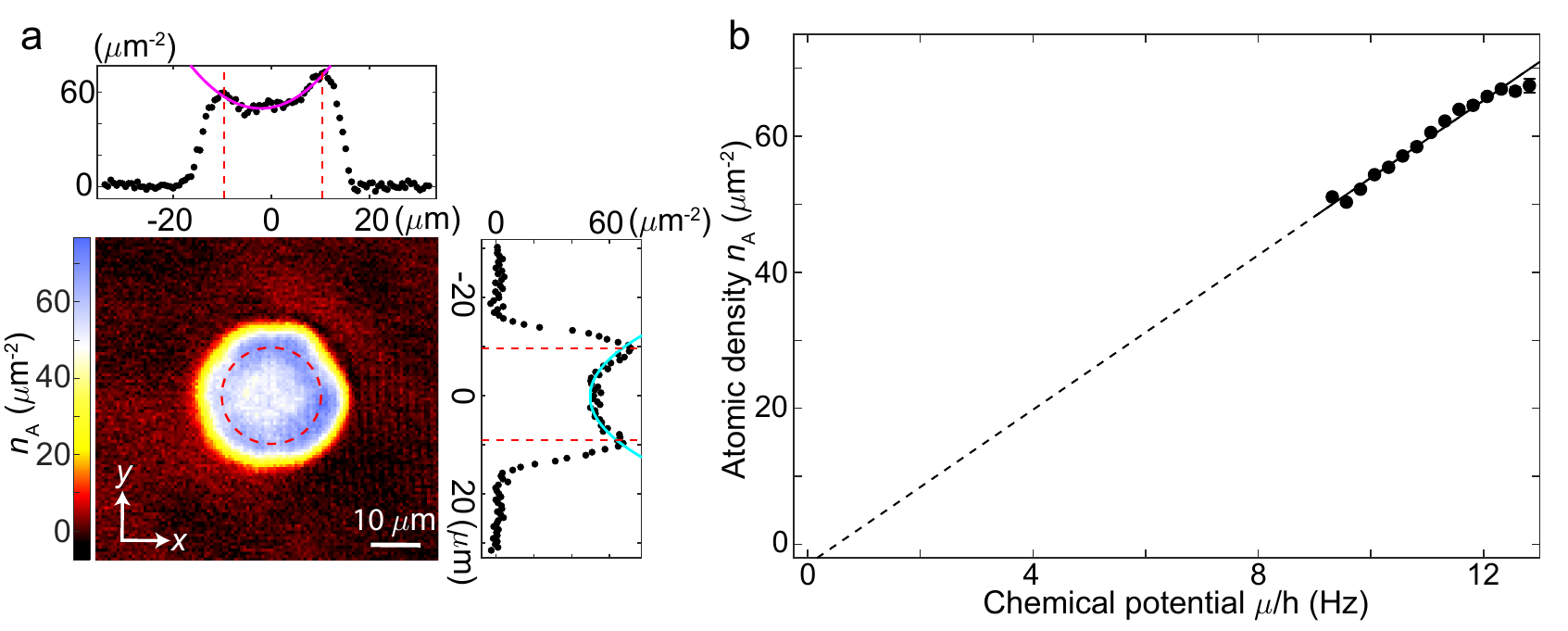}
    \caption{\textbf{Calibration of magnetic anti-trap potential from atomic density distribution.} \textbf{a}, Fit of the \textit{in situ} atomic density profile for determination of the magnetic anti-trap frequencies $\omega_x$ and $\omega_y$ using Eq.~\ref{nxy}. The top and right panels show line cuts of the 2D atomic density in x and y directions crossing at the center of the anti-trap. We choose the region within the red dashed circle for fit and extraction of the equation of state. \textbf{b}, Equation of state of atomic BEC shown in \textbf{a}. Each data point represents averaged density within a bin size $\delta\mu/h = 0.25$~Hz and error bars represent 1-$\sigma$ standard deviation. The black solid line is a linear fit to the data while the black dashed line is an extrapolation of the fit toward origin. }
    \label{fig:FigS2}
\end{figure}

The magnetic anti-trap frequency in the horizontal plane is given by
\begin{align}
    \omega_i^2 = \frac{\mu_m}{4mB_0}(B'^2-4\epsilon_iB_0^2),
    \label{TrapFrequency}
\end{align}
where $i = x,y$, $\mu_m$ is magnetic moment, $B_0$ and $B'$ are magnetic field and magnetic field gradient, respectively, at the location of particles and $\epsilon_i$ is determined from the coil geometry~\cite{CLthesis}. We determine the offset field value $B_0$ with an accuracy of 2~mG using microwave spectroscopy. We prepare atomic BEC at 17.2~G where the $s-$wave scattering length  $a_\mathrm{S}=4~a_0$. Because of the low chemical potential, the atomic density distribution is sensitive to the magnetic anti-trap and shows lower density at the center and higher density in the rim, see Fig.~\ref{fig:FigS2}a. Since the vertical trap frequency $\omega_z/2\pi = 400$~Hz is much larger than the chemical potential $\mu_0/h\approx 10$~Hz, the BEC is in the quasi-2D regime and the column density under Thomas-Fermi approximation is given by
\begin{align}
    n(x,y) = \frac{m}{\hbar^2g_{\mathrm{2D}}}[\mu_0 - V_{\mathrm{mag}}(x,y)],
    \label{nxy}
\end{align}
where the 2D coupling strength $g_{\mathrm{2D}} = \sqrt{8\pi}a_\mathrm{S}/\sqrt{\hbar/m\omega_z}$, the magnetic anti-trap potential $V_{\mathrm{mag}}(x,y)$ is parametrized by the trap frequencies $\omega_x$ and $\omega_y$ as $V_{\mathrm{mag}}(x,y) = m\omega_x^2(x-x_0)^2/2+m\omega_y^2(y-y_0)^2/2$ and $(x_0,y_0)$ is the center position of the anti-trap. To determine the trap frequencies and the global chemical potential, we fit the \textit{in situ} atomic density distribution using Eq.~\ref{nxy}, see Fig.~\ref{fig:FigS2}a. From the fit we get $\omega_x/2\pi = 1.94(9)$~Hz, $\omega_y/2\pi = 2.24(9)$~Hz and $\mu_0 = h\times 9.19(7)$~Hz. In this way, we calibrate the geometric parameters to be $\epsilon_x = 0.54(3)~\mathrm{cm}^{-2}$ and $\epsilon_y = 0.45(3)~\mathrm{cm}^{-2}$, which we use to calculate the anti-trap frequencies for molecules based on Eq.~\ref{TrapFrequency}, which gives $\omega_x^{mol}/2\pi = 3.35(4) $~Hz and $\omega_y^{mol}/2\pi = 3.48(4)$~Hz. For consistency check, we plot out the atomic density $n_{\mathrm{A}}$ versus the local chemical potential $\mu = \mu_0 - V_{\mathrm{mag}}(x,y)$, which agrees with the equation of state of a pure 2D BEC $\mu = (\hbar^2g_{\mathrm{2D}}/m)n(x,y)$ (see Fig.~\ref{fig:FigS2}b).

\begin{figure}
    \centering
    \includegraphics[width = 172mm]{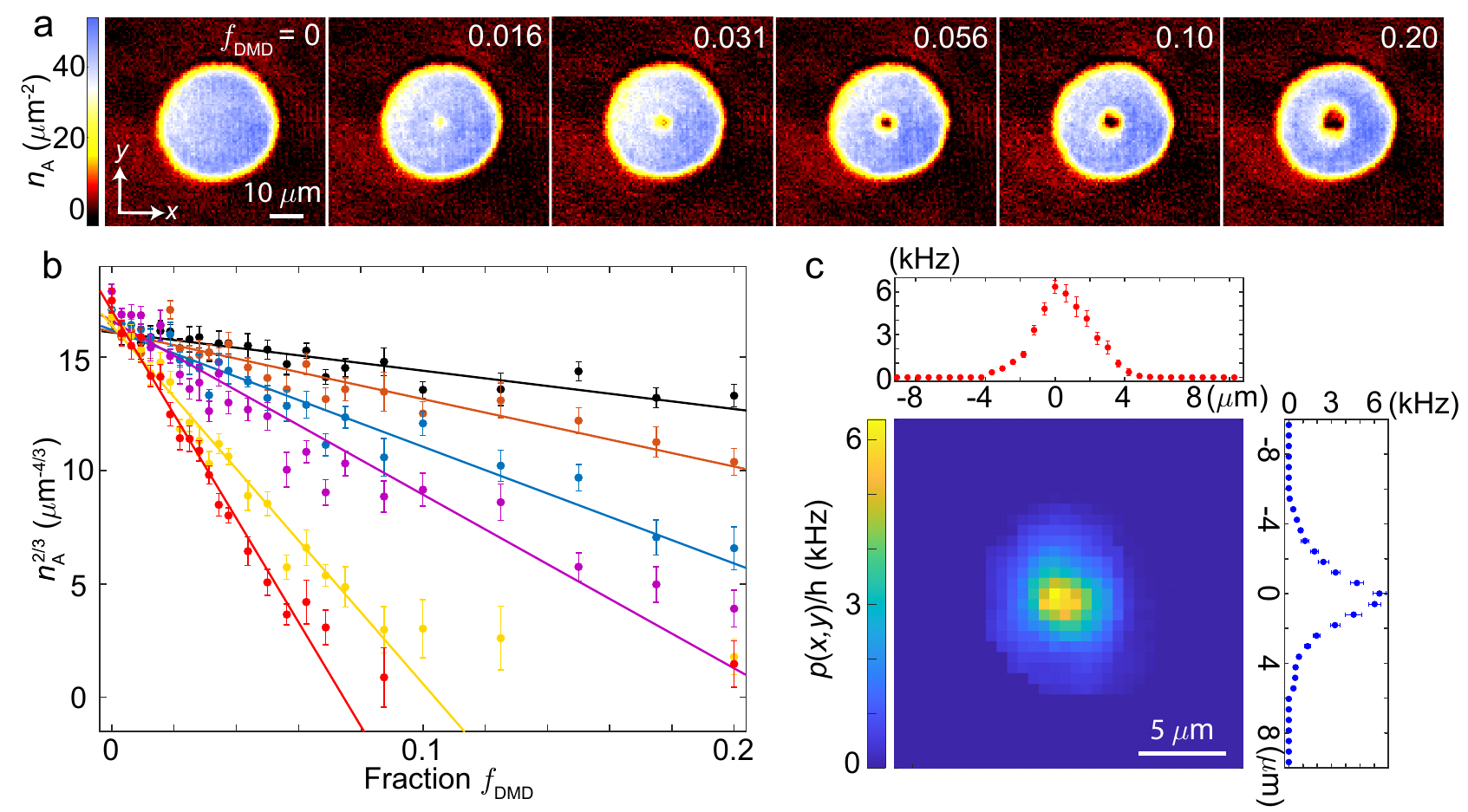}
    \caption{\textbf{Calibration of the optical potential barrier projected by DMD from the density response measurement of atomic BEC.} \textbf{a}, Images of \textit{in situ} atomic column density with different central barrier height determined by different fraction of micromirrors $f_{\mathrm{DMD}}$ that are turned on in DMD. \textbf{b}, Example measurements of the proportionality $p(x,y)$ for 6 pixels at different locations. The solid lines are linear fits to the linear part of the data points, the slope of which gives $p(x,y)$. \textbf{c}, Spatial dependence of the proportionality $p(x,y)$. The upper and right panels are line cuts in x and y directions crossing the peak value. }
    \label{fig:FigS3}
\end{figure}

We calibrate the optical potential barrier projected by DMD using atomic BEC prepared at 19.2~G, where the atomic  scattering length is $a_\mathrm{S} = 127$~$a_0$ and the vertical trap frequency is $\omega_z/2\pi = 409$~Hz. The intensity of the optical barrier is ramped up within 10~ms. After waiting for another 2~ms, absorption imaging is performed in the vertical direction to record the atomic column density, see Fig.~\ref{fig:FigS3}a. Here the barrier height is controlled by the fraction of micromirrors $f_{\mathrm{DMD}}$ that are turned on. The fraction determines the intensity of the light projected onto the atom plane. In the region with higher light intensity, the atomic density is suppressed more, which in turn allows us to determine the light intensity. Because of the higher chemical potential of BEC in this case, the density depletion has a larger dynamical range that helps to calibrate larger range of barrier height. Since the chemical potential is comparable to the vertical trap frequency, the BEC is in 3D regime and the column density under Thomas-Fermi approximation is given by
\begin{align}
    n(x,y) = [\alpha(\mu_0-V_{\mathrm{opt}}(x,y))]^{3/2}
    \label{EoS3D}
\end{align}
where $\alpha = (4\sqrt{2}/3g\sqrt{m}\omega_z)^{2/3}$, the 3D coupling strength $g = 4\pi\hbar^2a_\mathrm{S}/m$ and the local optical potential $V_{\mathrm{opt}}(x,y)$ is proportional to the micromirror fraction as $V_{\mathrm{opt}}(x,y) = p(x,y)f_{\mathrm{DMD}}$. Thus for each pixel located at (x,y), we have $n^{2/3}(x,y) = \alpha[\mu_0 - p(x,y)f_{\mathrm{DMD}}]$, from which the proportionality $p(x,y)$ can be extracted from a series of measurements with different $f_{\mathrm{DMD}}$, see Fig.~\ref{fig:FigS3}b. Repeating the same procedure for all the pixels within the region of optical barrier, we can map out the spatial dependence of the proportionality $p(x,y)$, see Fig.~\ref{fig:FigS3}c. The polarizability of weakly bound molecules is approximately twice as large as that of a free atom, thus the corresponding proportionality for the molecules is $2p(x,y)$.

After calibrating both the magnetic potential $V_{\mathrm{mag}}(x,y)$ and the optical potential $V_{\mathrm{opt}}(x,y)$ for molecules, we can get the local molecular density as a function of the total external potential $V(x,y) = V_{\mathrm{mag}}(x,y) + V_{\mathrm{opt}}(x,y)$ and follow the fitting procedure in Sec.~III to extract the global chemical potential $\mu_0$. Then we obtain the corresponding local chemical potential $\mu$ and average the density in a certain spatial area with a proper range of local chemical potential to get the equation of state for molecules from the density profile and optical barrier measurements in Fig.~\ref{fig:Fig2}. In addition, with the knowledge of the optical potential profile, we get the equation of state for the  BEC in 3D regime, see the inset of Fig.~\ref{fig:Fig2}c.

\section{Fitting the equation of state for 2D and 3D Bose gas}
For a nondegenerate 2D ideal Bose gas, the phase space density is given by $n_{\phi} = -\ln(1-\zeta)$, where $\zeta = \exp(\beta\mu)$ is the fugacity, $\beta = 1/k_{\mathrm{B}}T$ and $\mu = \mu_0 - V(x,y)$ is the local chemical potential. If the gas is interacting, a mean field potential $2(\hbar^2g_{\mathrm{2D}}/2m)n(x,y)$ is added to the external potential based on the Hartree-Fock approximation~\cite{Dalibard2011}. Then the equation of state for interacting 2D Bose gas becomes:
\begin{align}
    n(x,y) = -\frac{1}{\lambda_{\mathrm{dB}}^2}\ln[1-e^{\beta\mu-g_{\mathrm{2D}}n(x,y)\lambda_{\mathrm{dB}}^2/\pi}],
    \label{2DInteracting}
\end{align}
which we use for fitting the data from the optical barrier measurement in Fig.~\ref{fig:Fig2}c for density $n_{\mathrm{M}} < 1~\mathrm{\mu m}^{-2}$. On the other hand, the density of 2D superfluid outside the fluctuation region is~\cite{Svistunov2001}:
\begin{align}
    n(x,y) = \frac{2\pi\beta}{g_{\mathrm{2D}}\lambda_{\mathrm{dB}}^2}\mu + \frac{1}{\lambda_{\mathrm{dB}}^2}\ln[2n(x,y)\lambda_{\mathrm{dB}}^2g_{\mathrm{2D}}/\pi-2\beta\mu],
    \label{2DBKT}
\end{align}
which we use for fitting the data from the density profile measurement in Fig.~\ref{fig:Fig2}c for density $n_{\mathrm{M}} > 4~\mathrm{\mu m}^{-2}$.

We perform a global fit to the data points within the range $n_{\mathrm{M}} < 1~\mathrm{\mu m}^{-2}$ and $n_{\mathrm{M}} > 4~\mathrm{\mu m}^{-2}$ using Eq.~\ref{2DInteracting} and Eq.~\ref{2DBKT}, respectively, with temperature T, global chemical potential $\mu_0$ and 2D coupling constant $g_{\mathrm{2D}}$ as fitting parameters. Since the experimental condition drifted, the global chemical potential between the optical barrier and density profile measurements are different, and the chemical potential difference $\delta\mu$ is also set as an free parameter in the global fit. The fit gives $T = 11(1)$~nK, $g_{\mathrm{2D}} = 0.19(3)$ and the global chemical potential for the optical barrier and density profile measurements as $h\times 45(7)$~Hz and $h\times 61(7)$~Hz, respectively. We also performed independent fits to the data at low density $n_{\mathrm{M}} < 1~\mathrm{\mu m}^{-2}$ and high density $n_{\mathrm{M}} > 4~\mathrm{\mu m}^{-2}$. The resulting temperatures are $10(3)$~nK and $11(1)$~nK, in agreement with each other and with the global fit.

With the extracted 2D coupling constant $g_{\mathrm{2D}}$, the critical phase density for BKT superfluid transition is evaluated as $\ln(\xi/g_{\mathrm{2D}})\approx 7.5$, where the coefficient $\xi = 380(3)$~\cite{Svistunov2001}. On the other hand, the BEC transition in our 2D box potential occurs at critical phase space density of $\ln(4\pi R^2/\lambda_{\mathrm{dB}}^2)\approx 7.5$~\cite{Dalibard2011}, which coincides with the BKT transition.

For BECs in 3D regime as shown in the inset of Fig.~\ref{fig:Fig2}c, the low density part where the column density $n_{\mathrm{A}} < 10~\mathrm{\mu m}^{-2}$ is fitted using the classical gas formula $n(x,y) = (2\pi l_z^2/\lambda_{\mathrm{dB}}^4)\exp(\beta\mu)$, where the harmonic oscillator length $l_z = \sqrt{\hbar/m\omega_z}$. The high density part is fitted based on Eq.~\ref{EoS3D}.

\section{Extraction of the two-body inelastic loss coefficients}
In order to study the lifetime of g-wave molecules, we hold the molecules in different traps and monitor the decay of particle number as a function of the hold time. The two traps we used have horizontal radius $R_1 = 12.5~\mathrm{\mu m}$, $R_2 = 9~\mathrm{\mu m}$ and vertical trap frequency $\omega_{z1}/2\pi = 400$~Hz, $\omega_{z2}/2\pi = 167$~Hz, respectively. The molecular density distribution in these traps are approximately uniform in the horizontal direction and Gaussian in the vertical direction, given by
\begin{align}
    n(\vec{r}) = \frac{N_{\mathrm{M}}}{\pi^{3/2} R_i^2l_{zi}}e^{-z^2/l_{zi}^2}\theta(R_i - \rho),
    \label{3D_density}
\end{align}
where $i$ = 1,2, $\rho = \sqrt{x^2+y^2}$ and $\theta(x)$ is the heaviside step function.

Even though the 1064~nm light intensity in the vertical direction of the two traps differ by a factor of $\omega^2_{z1}/\omega^2_{z2}\approx 6$, the decay rate of molecular number are similar, see Fig.~\ref{fig:Fig3}a. This suggests that the one-body loss process due to the off-resonant laser light is negligible. In fact, since the g-wave molecules are in a highly excited rovibrational state, two-body loss process dominates which is modelled by $\partial_t n(\vec{r},t) = -L_2n^2(\vec{r},t)$. The molecular number decay corresponding to the density profile in Eq.~\ref{3D_density} is thus given by
\begin{align}
    N_{\mathrm{M}}(t) = \frac{N_{\mathrm{M}}(0)}{1+L'_2N_{\mathrm{M}}(0)t},
    \label{N_Mt}
\end{align}
where $L'_2 = L_2/\sqrt{2}\pi^{3/2}R_i^2l_{zi}$. We use Eq~\ref{N_Mt} to fit the data of molecular number decay and extract the inelastic loss coefficient $L_2$ in Fig.~\ref{fig:Fig3}b.

The unitarity limit of the two-body loss coefficient is $U_2(k) = 4h/2mk$, where $k$ is the magnitude of the relative wavevector $\vec{k}$ between two colliding molecules, associated with the relative kinetic energy $E = \hbar^2\vec{k}^2/2m$~\cite{Chin2010}. Due to the finite temperature in our experiment, the relative kinetic energy obeys Bolzmann distribution as $p(E) = A\exp(-E/k_{\mathrm{B}}T)$, where the coefficient $A = (1/4)(\hbar^2/\pi mk_{\mathrm{B}}T)^{3/2}$. The distrbution of the wavenumber $k$ is then given by

\begin{align}
    p(k) = 4\pi Ak^2e^{-\hbar^2k^2/2mk_{\mathrm{B}}T}.
    \label{p_of_k}
\end{align}
The unitarity limit that we evaluate in Fig.~\ref{fig:Fig3}b is $U_2 = \int_0^{\infty}U_2(k)p(k)dk = (4h/2m)\langle k^{-1}\rangle$, where the thermal average of $k^{-1}$ with respect to the distribution $p(k)$ is $\langle k^{-1}\rangle = \sqrt{\hbar^2/\pi mk_{\mathrm{B}}T}$. For comparison with the loss coefficients, we evaluate the interaction scale as $\mu_0/\hbar n_{\mathrm{3D}}$, where the 3D mean density $n_{\mathrm{3D}} = \int_{-\infty}^{+\infty}n^2(\vec{r})d^3\vec{r}/\int_{-\infty}^{+\infty}n(\vec{r})d^3\vec{r} = N_{\mathrm{M}}/\sqrt{2}\pi^{3/2}R_1^2l_{z1}^2$.

\section{Empirical fit to dissociation rate and dissociated molecular fraction}
After preparing a pure molecular BEC below the Feshbach resonance, if the magnetic field is then switched to a value high above the resonance, the molecules quickly dissociate into a continuum of free atoms. The dissociation rate follows Fermi's golden rule as $\Gamma = (2\pi/\hbar)|V_{\mathrm{MA}}|^2\rho(E) = 2m^{1/2}a_{\mathrm{bg}}\Delta\mu\Delta BE^{1/2}/\hbar^2$, where $V_{\mathrm{MA}}$ is the coupling matrix element between molecular and atomic states and is independent of the energy E above the continuum to leading order, the density of state $\rho(E)\propto E^{1/2}$ and $a_{\mathrm{bg}}$ is the background scattering length. In this high field limit, our measured dissociation rate is consistent with Fermi's golden rule $\gamma = \alpha\Gamma$, where the coefficient $\alpha = 0.4(1)$. The dissociation rate in Fig.~\ref{fig:Fig4}b is extracted by fitting the data in Fig.~\ref{fig:Fig4}a using the formula $N_{\mathrm{M}}(t) = N_{\mathrm{M}}(t_0)\{1-\exp[-\gamma (t-t_0)]\}\theta(t-t_0)$, where $t_0$ is the time when the molecules start to dissociate. 

On the other hand, when the magnetic field is ramped to near the resonance where $\rho(E)\approx 0$, we still observe a finite dissociation rate of $8~\mathrm{ms}^{-1}$. This is because the molecular state can couple to a band of scattering states that are Lorentzian distributed~\cite{Fano1961}. We thus define an effective density of state $\rho_{\mathrm{eff}}$ to be a convolution between $\rho(E)$ and a Lorentzian distribution. Thus the effective dissociation rate becomes
\begin{align}
    \Gamma_{\mathrm{eff}} = \Gamma\sqrt{(\sqrt{1+\Omega^2/4E^2}+1)/2},
    \label{Gamma_eff}
\end{align}
where $\Omega$ is the full width of the Lorentzian distribution. We use Eq.~\ref{Gamma_eff} to fit the dissociation rate as a function of the magnetic field we measured, shown as the blue solid line in the upper panel of Fig.~\ref{fig:Fig4}b.

The dissociated molecular fraction drops when the magnetic field is ramped back closer to the resonance, which we attribute to the inelastic collision loss between atoms and molecules near the resonance. The data of the fraction in Fig.~\ref{fig:Fig4}b is fitted using an empirical function $f = N_s\{1/2 + (1/\pi)\arctan[\Delta\mu(B-B_0)/V_s]\}$, where $N_s$ and $V_s$ are set as free parameters.

\end{document}